\documentclass[sigconf,9pt]{acmart}
\AtBeginDocument{%
  \providecommand\BibTeX{{%
    \normalfont B\kern-0.5em{\scshape i\kern-0.25em b}\kern-0.8em\TeX}}}

\setcopyright{acmcopyright}
\copyrightyear{2019}
\acmYear{2019}

\acmConference[AIChallengeIoT'19]{First International Workshop on Challenges in Artificial Intelligence and Machine Learning}{November 10--13, 2019}{New York, NY, USA}
\acmBooktitle{First International Workshop on Challenges in Artificial Intelligence and Machine Learning (AIChallengeIoT'19), November 10--13, 2019, New York, NY, USA}
\acmDOI{10.1145/3363347.3363366}

\usepackage{multirow}

\usepackage{tikz}
\usetikzlibrary{arrows,positioning} 

\usepackage{amsmath}

\usepackage{algpseudocode}
\usepackage{algorithm}
\usepackage{soul}

\usepackage[caption=false]{subfig}

\usepackage{array}
\usepackage{amssymb}
\usepackage{hyperref}
\usepackage{multicol}
\usepackage{caption}

\usepackage{booktabs}
\usepackage{amsthm}
\theoremstyle{definition}

\newtheorem{definition}{Definition}

\def\schemeNameRaw{deepFogGuard}
\def\schemeName{{\it \schemeNameRaw}}

\def\camera{multi-camera object classification}
\def\health{health activity classification}

\def\low{{\em Hazardous}}
\def\med{{\em Poor}}
\def\high{{\em Normal}}
\def\nofailure{{\em No Failure}}

\begin{document}

\title{{\em Guardians of the Deep Fog}: Failure-Resilient DNN Inference from Edge to Cloud}


\settopmatter{authorsperrow=4}

\author{Ashkan~Yousefpour}
\authornote{These authors contributed equally to the paper.}
\authornote{ashkan.y@berkeley.edu}
\affiliation{
  \institution{UT Dallas $\&$ UC Berkeley}
}

\author{Siddartha~Devic}
 \authornotemark[1]
\affiliation{
  \institution{UT Dallas}
}

\author{Brian~Q.~Nguyen}
\authornotemark[1]
\affiliation{
  \institution{UT Dallas}
}

\author{Aboudy~Kreidieh}
\affiliation{%
  \institution{UC Berkeley}
}

\author{Alan~Liao}
\affiliation{%
  \institution{UT Dallas}
}

\author{Alexandre~M.~Bayen}
\affiliation{%
  \institution{UC Berkeley}
}

\author{Jason~P.~Jue}
\affiliation{%
  \institution{UT Dallas}
}

\renewcommand{\shortauthors}{A. Yousefpour, et al.}

\begin{abstract}
  Partitioning and distributing {\em deep neural networks} (DNNs) over physical nodes such as edge, fog, or cloud nodes, could enhance sensor fusion, and reduce bandwidth and inference latency. However, when a DNN is distributed over physical nodes, failure of the physical nodes causes the failure of the DNN units that are placed on these nodes. The performance of the inference task will be unpredictable, and most likely, poor, if the distributed DNN is not specifically designed and properly trained for failures. Motivated by this, we introduce \schemeName{}, a DNN architecture augmentation scheme for making the distributed DNN inference task failure-resilient. To articulate \schemeName{}, we introduce the elements and a model for the resiliency of distributed DNN inference. Inspired by the concept of residual connections in DNNs, we introduce skip hyperconnections in distributed DNNs, which are the basis of \schemeName{}'s design to provide resiliency. Next, our extensive experiments using two existing datasets for the sensing and vision applications confirm the ability of \schemeName{} to provide resiliency for distributed DNNs in edge-cloud networks. 
\end{abstract}

%



\maketitle

\section{Introduction}
With the proliferation of the {\em Internet of Things} (IoT) applications, increasing numbers of smart IoT devices are being deployed and integrated into our daily routines. Smart homes, smart cities, wearables, self-driving vehicles, AR and VR, context sensing and crowdsensing, and smart retail are examples of adaptations of IoT devices into human spaces~\cite{Deepiot,fog-survey}. To intelligently analyze and act on the data that IoT devices generate, {\em machine learning} (ML) techniques are seen to be promising. This is primarily because IoT devices are often directly connected to data sources, such as cameras, microphones, gyroscopes, or sensors that capture a large quantity of input data that could feed the ML models~\cite{dnnApplication,harvard}. 

Due to their accuracy and powerful expressiveness, deep learning methods, among other ML techniques, have been a successful choice for IoT applications in a broad spectrum of domains such as computer vision, speech recognition, medical diagnosis, and natural language processing~\cite{park2018wireless,Li:Edge}. Deep learning techniques make use of {\em deep neural networks} (DNNs). In certain DNN-empowered IoT applications, the {\em inference} task runs for a prolonged period of time. Examples of such IoT applications are image-based defect detection in a factory, automatic recognition of parts during product assembly, or anomaly behavior detection in a crowd based on DNNs \cite{chen2019exploring}. Nevertheless, a challenge with DNN-empowered IoT applications is determining where the DNN model should be placed for the inference task.

The immediate option may be deploying DNNs directly onto the IoT devices; however, this is often infeasible, as many IoT devices are resource-constrained and cannot efficiently support the computational requirements of DNNs. For instance, according to Liu {\em et al.} \cite{EdgeEye}, the GoogleNet model for image classification is larger than 20 MB and requires about 1.5 billion multiply-add operations per inference per image. 

Another possibility is to place the DNN in the cloud and send the IoT data to the cloud, since the cloud servers are equipped with powerful hardware such as TPUs and GPUs. Nevertheless, when a DNN is deployed in the cloud, the data has to be continuously transmitted from IoT devices to the cloud in WAN environments during inference, which results in the heavy consumption of network resources, high latency, and privacy concerns~\cite{harvard, EdgeEye}. 

Another option for DNN placement is to distribute the DNN over physical nodes along an edge-fog-cloud hierarchy~\cite{harvard, morshed2017deep, EdgeEye, hotedge-distributed, ChuangHu, dey1}. The idea of thie current approach is to distribute the DNN onto {\em edge nodes}, {\em fog nodes}, and {\em cloud nodes} so that inference from IoT data is processed along the route, on different physical nodes from the edge to cloud.

A natural question that arises with this approach is whether the resulting distributed DNN inference along edge-fog-cloud is resilient in the presence of physical nodes failures. Specifically, the question is: {\em what happens to an ongoing inference task of a distributed DNN when its physical nodes fail, and how can we make distributed DNN inference resilient to physical node failures?} This question is the topic of our study. 

When a DNN is distributed over physical nodes, failure of a physical node causes the failure of the DNN units that are placed on the node. Failure of physical nodes could be due to power outages, cable cuts, natural disasters, or hardware/software failures. The effect of such failures on distributed DNN inference may be heavily dependent on the time to recover from the failures. 

While the physical nodes are being recovered, the performance of the distributed DNN inference is unpredictable, and most likely, poor, if the distributed DNN is not specifically designed and properly trained for failure resiliency. This is especially important for critical applications that cannot tolerate unpredictable and poor performance, even for a short time. 


In this article, we study the failure resiliency of distributed DNN inference over the edge, fog, and cloud nodes, where the failure of a physical node results in the failure of the DNN units that are placed on the node. Our main contributions in this article are 
\begin{enumerate}
    \item {\em We introduce \schemeName{}, a DNN architecture augmentation scheme for making the distributed DNN inference failure-resilient}: In order to provide context for \schemeName{}, we introduce the elements of distributed DNNs and  \schemeName{}. Inspired by the concept of residual connections in DNNs~\cite{ms-residual}, we introduce \textbf{skip hyperconnections} in distributed DNNs, which are the basis of \schemeName{}'s design to provide resiliency. Residual connections skip one or more DNN layers, whereas skip hyperconnections skip one or more physical nodes. 
    \item {\em We conduct extensive experiments using two existing data sets for sensing and vision applications}: We construct a model for measuring the resiliency of inference in distributed DNNs. Finally, we confirm the ability of \schemeName{} to provide resiliency for distributed DNNs in edge-cloud networks.
\end{enumerate}
In \schemeName{}, upon failure of a physical node, the information flow can still be routed through the distributed DNN, thanks to the skip hyperconnections. Hence, we call the skip hyperconnections the {\em Guardians of the Deep Fog}, since they act as the guard of information flow in distributed DNNs over edge-fog-cloud hierarchy.


\section{Definitions} \label{definitions}
In this section we introduce the definitions required to articulate \schemeName{} and provide the necessary context. 







{\bf Partitioning and distributing DNNs}. $\textreferencemark$ A deep neural network $G$ can be split according to a partition map $u$ and can be distributed over a set of physical nodes $V$. We denote the partition operation by $\oslash$ and the resulting split DNN by $G \oslash^u V$. The present study does not address the problem of optimal partitioning of the DNNs (i.e. finding an optimal partition map $u$), as it is not the primary focus of this article. The optimal DNN partitioning is non-trivial and depends on many factors including available network bandwidth, type of DNN layers (convolutional vs. fully-connected), and DNN graph topology~\cite{neurosurgeon, wangplacement, dey1, zhou2019distributing, ChuangHu, Li:Edge, elgamal2018droplet}. 
Instead, this article studies the resiliency of {\em previously-partitioned} distributed DNN models during inference.

{\bf Previously partitioned DNN}. If $u_*$ denotes the desired partition map for a certain use case (with regards to constraints such as delay, bandwidth, or energy), $G_V$ denotes the resulting partitioned DNN according to $u_*$. Hence, $G_V=G \oslash^{u_*} V$.

{\bf Physical Nodes vs. DNN Units}: To distinguish between physical nodes in the network and artificial nodes (neurons or units) in DNNs, we clarify by using {\em units} when referring to DNN neurons. Additionally, we only use the term {\em physical node}; however, the concepts in this study are also applicable to {\em virtual nodes}, such as VMs and containers.

{\bf Types of Physical Nodes}: IoT devices (e.g. sensors, cameras, or mobile phones) are usually the main sources of data, whereas cloud servers are central hubs for processing and storage. Cloud servers are normally part of large data centers. On the other hand, fog nodes could host services packaged in the form of VMs, containers, or unikernels, and can be routers, switches, dedicated servers for fog computing (e.g. {\em cloudlets}), set-top boxes, access points, or firewalls. Similarly, edge nodes are devices attached to the connected things, such as WiFi access points, first-hop routers and switches, and base stations \cite{fog-survey}.

\begin{figure}[!t]
\centering
    \includegraphics[width=0.85\linewidth]{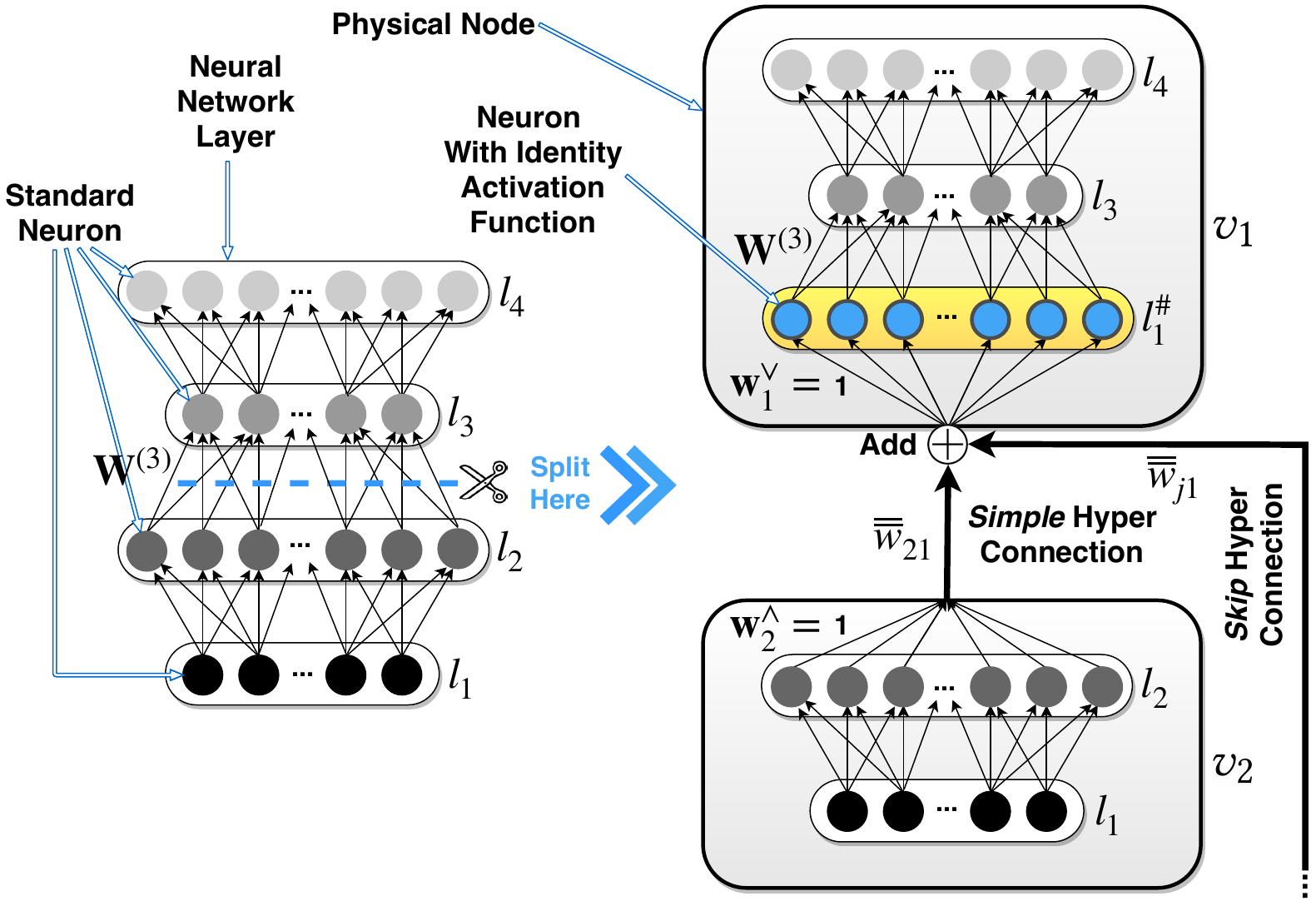}
    \caption{Partitioning a DNN and distributing it across physical nodes $v_1$ and $v_2$. The DNN is fully connected (not all weights are shown).}
    \label{fig:split}
\end{figure}

Figure~\ref{fig:split} shows the process of splitting a DNN with four fully-connected layers and distributing it across two physical nodes $v_1$ and $v_2$. The layers $l_1$ and $l_2$ are stored on node $v_2$ and the layers $l_3$ and $l_4$ are stored on node $v_1$. $\mathbf{W}^{(3)}$ is matrix of weights at layer $l_3$. Note that a special layer $l^{\#}_1$ (called {\em expansion layer}) is included below layer $l_3$ in node $v_1$. This layer is for the reception and the expansion of the vector of data from the other physical nodes (to be discussed). Since the distributed DNN resides on different physical nodes, during inference the vector of output values from one physical node must be transferred (e.g. through a TCP socket) to another physical node. We call the transfer link (pipe) between two physical nodes a {\em hyperconnection}. 

{\bf Hyperconnections}: Unlike a typical neural network that connects two units and transfers a scalar, a {\em hyperconnection} connects two physical nodes over which the layers of a DNN is distributed and transfers a vector of scalars. 

{\bf Simple vs. Skip Hyperconnections}:
Hyperconnections are one of two kinds: {\em simple} or {\em skip}. A hyperconnection is called {\em simple} when it connects a physical node to the physical node that has the next DNN layer (i.e. ``parent'' node in the hierarchy), and is called {\em skip} when it skips one or more physical nodes in the hierarchy and connects a physical node to an ``ancestor'' node. In Fig.~\ref{fig:split}, the simple hyperconnection between $v_1$ and $v_2$ connects the output values of layer $l_2$ to the input of expansion layer $l^{\#}_1$. The skip hyperconnection connects the output of a ``descendant'' physical node $v_j$ (not shown) to the input of expansion layer $l^{\#}_1$. 

The concept of skip hyperconnection is similar to that of residual connections in DNNs~\cite{ms-residual}. Residual connections are a special case of highway connections~\cite{highway, ms-residual} and are those connections skipping one or more DNN layers. Similarly, skip hyperconnections skip one or more physical nodes in a distributed DNN. \schemeName{} makes use of skip hyperconnections for added resiliency, so that upon failure of a physical node, the information flow can still be routed to the cloud for inference. In Section~\ref{model} we explain that adding skip hyperconnections improves resiliency of the distributed DNN. 

{\bf Remark}: Note that the weights that feed the unit output values of physical node $v_2$ to the hyperconnection (vector $\mathbf{w}^{\wedge}_2$) and the weights that expand the output of the {\em Add} operation to the expansion layer of physical node $v_1$ (vector $\mathbf{w}_1^{\vee}$) are all set to 1. The value of these vectors can be chosen arbitrarily because distributed DNNs learn to adjust and compensate their weights during training (the training process will be discussed soon). For simplicity but without loss of generality, we assume the value of $\mathbf{1}$ for these vectors.

{\bf Hyperconnection Weights}: Similar to connection weights in neural networks, hyperconnections may also have a {\em weight} vector; the elements of the vector that passes through the hyperconnection are multiplied by this weight. Let vector $\mathbf{\overline{\overline{w}}}_{ij}$ denote the hyperconnection weight that connects physical node $v_i$ to physical node $v_j$. In this study, $\forall i,j~~ \mathbf{\overline{\overline{w}}}_{ij}=\mathbf{1}$, that is the weight of all hyperconnections is chosen to be the vector $\mathbf{1}$ (a vector with all elements equal to 1).

\begin{definition}
{\em Adding hyperconnections' inputs.} The {\em Add} $\bigoplus$ operation is an element-wise vector addition that adds the elements of two or more hyperconnections. When $H_i$ denotes the set of indices of all physical node that have a hyperconnection to the physical node $v_i$, the Add operation at node $v_i$ computes the vector $\mathbf{x}^{(l_i^\#)}$, the input vector to the expansion layer of node $v_i$, by adding the data going through the incoming hyperconnections as 
\begin{equation}
    \mathbf{x}^{(l_i^\#)}=\sum_{j\in H_i}{\mathbf{\overline{\overline{w}}}_{ji}\mathbf{x}_{ji}},
\end{equation}
where $\mathbf{x}_{ji}$ is the vector passing through the hyperconnection connecting the physical node $v_j$ to the physical node $v_i$. Since, in this article, $\mathbf{\overline{\overline{w}}}_{ij}=\mathbf{1}$, we will have $ \mathbf{x}^{(l_i^\#)}=\sum_{j\in H_i}{\mathbf{x}_{ji}}$. 
\end{definition}

We represent the vector output of a failed physical node {i.e. \em null vector}, by the symbol $\Phi$. Formally, when a null vector is added to a non-null vector, the null vector is ignored. That is, $\mathbf{x}_{ij}\bigoplus\Phi=\mathbf{x}_{ij}$. In the case where all source physical nodes of the incoming hyperconnections to a physical node fail, we will have $\Phi\bigoplus\ldots\bigoplus\Phi=\Phi$, which means the input of the physical node will be the null vector. Since in this case, applying operations on the null vector is meaningless, the physical node outputs the null vector $\Phi$. If the null vector is propagated all the way through the last layer of DNN, which means the information flow did not make it to the last layer, random guessing is performed.

{\bf Hyperconnection Dimensions}: The Add operation requires that the dimensions of the operands be the same. However, the dimensions of the hyperconnections are not always the same, as they connect layers of DNN in different depths. When it is necessary to change the dimension of the vectors, we perform zero-padding to match the vector with the largest dimension. The vector with the largest dimension also decides the dimension (number of units) of the {\em expansion layer}. 

{\bf Expansion Layer}: A special layer called the {\em expansion layer} is added to a physical node for the reception of the input vector from the Add operation ($\mathbf{x}^{(l_i^\#)}$) and its expansion to the units of the first layer at the physical node. The output (vector) of the Add operation must be connected to the corresponding units in the expansion layer. Hence, each unit in the expansion layer has one input, the corresponding value in the output vector. The units in the expansion layer all have identity function as their activation function, so that they do not change the incoming data. The number of the units in the expansion layer should be equal to the number of elements in the output vector of the Add operation, which is the maximum among the dimensions of the added vectors going through hyperconnections.

{\bf Training Process}: Once the elements of the distributed DNN (hyperconnections and expansion layers) are added, the distributed DNN must be trained with the new elements. The training process need not to be distributed, that is, when the DNN is physically distributed; the training process can be executed in a ``simulated'' environment, where the DNN is distributed in a simulation.  

\section{\schemeNameRaw{} Design} \label{model}
Distributed DNNs cannot be considered intrinsically failure-resilient without a proper design. Conventionally, we refer to the distributed DNNs that are not trained for failure resiliency as {\em Vanilla}. 

\schemeName{}'s goal is to increase the resiliency of the distributed DNN. We tend towards {\em passive resiliency}, the ability to function in the presence of failure without any re-training or reacting, but by exploiting the intrinsic resiliency \cite{torres2017fault}. To accomplish this, we consider and experiment with a method that augments the training process with built-in resiliency: adding skip hyperconnections to the architecture of a distributed DNN. Skip hyperconnections inherently increase the resiliency of the underlying neural architecture. 

{\bf Resiliency via Skip Hyperconnections}: The concept of skip hyperconnections is similar to residual connections. DNNs with residual connections are easier to optimize and have been shown to implicitly deal with the exploding gradient problem, ultimately providing better performance than standard DNNs \cite{ms-residual}. Inspired by residual connections in DNNs, in \schemeName{} we add skip hyperconnection between physical nodes to provide additional pathways for the flow of information through the model, even in the presence of partial failures. This is to thwart the {\em no-information-flow} situation, in which the information does not make it to the last layer of the DNN, and random guessing has to be performed.  

Figure~\ref{fig:experiment_setup} depicts the architecture of our experiments, discussed in Section~\ref{experiment}. The dashed arrows represent skip hyperconnections between physical nodes. (The expansion layers are not shown in Fig.~\ref{fig:experiment_setup}, since the layers have the same dimension in each experiment.) In the architecture on the right in Fig.~\ref{fig:experiment_setup}, all seven skip hyperconnections that skip one physical node are present. Similarly, all the three skip hyperconnections that skip one physical node are present in the distributed DNN on the left in Fig.~\ref{fig:experiment_setup}. Generally, it is expected that more skip hyperconnections improve resiliency, especially in more extreme failure scenarios. Nevertheless, we found that the determining the number of skip hyperconnections is a non-trivial task and depends on the learning task, reliability setting, and original DNN architecture. Determining the right skip hyperconnections could be done during training. Resource heterogeneity across edge, fog, and cloud nodes could also be a deciding factor for setting up the skip hyperconnections.

{\bf Implementation Notes}: (1) Skip hyperconnection can be implemented via TCP connections. (2) When implementing the Add operation, one has to ensure that the failure of a physical node does not result in an interruption, e.g. when a {\tt TCP socket exception} is thrown. (3) When a physical node fails, another physical node should be able to ``sense'' the failure. Physical nodes are responsible for checking the hyperconnections' respective source nodes from which they are fed. This can be done through a simple {\em keep-alive} mechanism. (4) Inference should be implemented in a synchronized fashion, that is, if a physical node fails and its output is not present, the null vector $\Phi$ should be used as the data. 

\section{Experiments} \label{experiment}
We need to construct a metric for measuring the resiliency of distributed DNNs. In the following subsection, we define a metric based on average accuracy to model resiliency of distributed DNNs. 

\subsection{Modeling Resiliency for Distributed DNN}

We consider a set of $n$ physical nodes $V = \{v_1,v_2,\ldots,v_n\}$ over which a DNN is distributed according to the partition map $u$. 

\begin{definition}
 {\em Reliability setting} of the physical nodes in $V$, is an $n$-tuple $R_V=(r_1,r_2,\ldots,r_n)\in [0,1]^n$, where each element $r_i$, referred to as {\em reliability probability}, is the probability that the physical node $v_i\in V$ survives at inference time. $r_i=(1-p_i)$, where $p_i\in [0,1] $ is the probability that the physical node $v_i\in V$ fails during inference time.   
\end{definition}

For the sake of easier terminology and notation, we utilize the convention in network reliability engineering, and use {\em reliability} over {\em probability of failure}. In order to model the reliability of a given distributed DNN, we need to model the following: (I) the physical nodes failing simultaneously during the inference time; and (II) the probability of a given simultaneous failure. To model (I), we introduce {\em node failure combination}, which is the combination of the physical nodes that fail simultaneously during the inference time.

\begin{definition}
A {\em node failure combination} of the physical nodes in $V$, is an $n$-tuple $B_V=(b_1,b_2,\ldots,b_n)\in\{0,1\}^n$, in which each element $b_i$ is a binary value indicating whether the physical node $v_i\in V$ has failed (0) or not (1). 
\end{definition}

We now have a model for (I), through $B_V$. Since, at inference time, a physical node $v_i\in V$ can either fail or survive and since $|V|=n$, there are $2^n$ possible node failure combinations (multiple physical nodes can fail at the same time). 

To model (II), we introduce $p(B_V|R_V)$, the probability of occurrence of a certain node failure combination $B_V$ during inference time, given a reliability setting $R_V$ over physical nodes $V$. This probability can be calculated as:

\begin{equation} \label{probability}
    p(B_V|R_V)=\prod_{i=1}^{|V|} {\big[b_ir_i + (1-b_i)(1-r_i)\big]}.
\end{equation}

As a numerical example, assume that a network has four physical nodes and reliability setting is $R_V=(0.98,0.98,0.95,0.94)$. The probability of the node failure combination where node 4 fails ($B_V=(1,1,1,0)$) is $p(B_V|R_V)=0.98\times0.98\times0.95\times0.06$.

Now we have a model for reliability of distributed DNNs. Next, we need to define a new notation $\varnothing$, necessary to model the resiliency of a distributed DNN:

\begin{definition}
{\em Operator $\varnothing$}. Given a node failure combination $B_V$, the notation $G_V\varnothing B_V$ represents the distributed DNN $G_V$ (over physical nodes $V$), in which those units that reside on $B_V$'s failing physical nodes are failed. 
\end{definition}

{\bf Average Accuracy as Resiliency Measure}: For modeling the resiliency of distributed DNNs, performance indicators such as accuracy, precision, or recall may be used. In this article, we use {\em average accuracy} as the measure of resiliency. 

\begin{figure}[t]
    \centering
    \includegraphics[width=\linewidth]{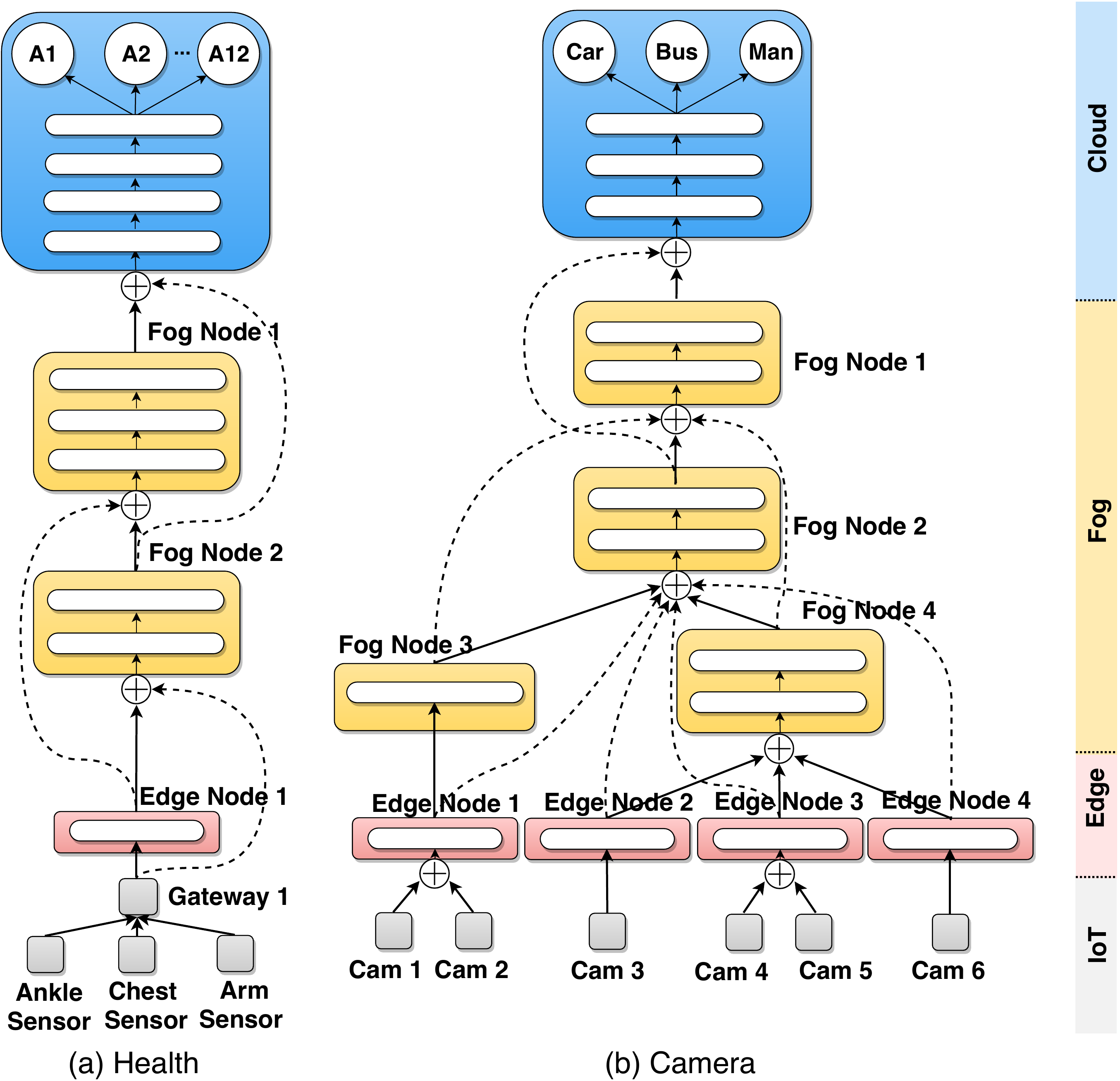}
    \caption{Distributed DNN architecture in \schemeName{} for: (a) \health{}, (b) \camera{} experiments. The expansion layers are not shown. (Physical node numbers are unique within each experiment)}
    \label{fig:experiment_setup}
\end{figure}

The average accuracy of a distributed DNN $G_V$ over physical nodes $V$ against reliability setting $R_V$ during inference is
\begin{equation} \label{resiliency}
    \overline{\mathbb{A}}(G_V, R_V) = \sum_{B_V} p(B_V|R_V)\times\mathcal{A}(G_V \varnothing B_V),
\end{equation}
where $\mathcal{A}(G_V \varnothing B_V)$ is the {\em accuracy} of $G_V \varnothing B_V$ during inference, and $p(B_V|R_V)$ is the probability of node failure combination $B_V$ given a reliability setting $R_V$. Equation (\ref{resiliency}) calculates the weighted average of accuracy over all possible node failure combinations (weighed by $p(B_V|R_V)$). Now we have a model for measuring resiliency of distributed DNNs. Next, we discus the datasets, the setup, and the results of our extensive experiment. 

\subsection{Datasets}
Our extensive experiments are conducted using two existing datasets for the sensing and vision applications, explained as follows.

{\bf Health Activity Classification (``Health'')}: We utilize the mobile health activity sensor dataset (UCI {\em MHealth}~\cite{mhealth}) as a benchmark for failure resiliency of a vertically distributed DNN (see Fig.~\ref{fig:experiment_setup}a). The dataset is comprised of readings from various sensors placed on different body parts of patients. This dataset is an example of an IoT application for medical purposes. The dataset contains sensor acceleration data from three different sensors placed at the chest, left ankle, and right arm. Additionally, the left ankle and right arm sensor provide body orientation data, and the chest sensors provide ECG measurements. The dataset is labeled with the 12 activities a patient is performing at a given time, and the task is to classify the type of activity (e.g. if the patient is sitting or running). There are a total of 23 features, where each feature corresponds to a specific type of data collected from one of the three sensors across ten human test subjects. For this experiment, the activities that do not belong to one of the 12 classes are removed, resulting in a dataset of 343,185 data points.  The \health{} dataset is approximately uniformly distributed across each class after preprocessing, and hence we use a standard cross-entropy loss function for the classification.

{\bf Multi-Camera Object Classification (``Camera'')}: The multi-view object detection dataset~\cite{multiview-dataset} is used as a benchmark for failure resiliency in DNNs that are distributed both vertically and horizontally (see Fig.~\ref{fig:experiment_setup}b). The dataset contains videos of a street from six different viewpoints, with object bounding boxes for frames captured from each camera. The bounding boxes are placed around three classes of objects: pedestrians, cars, and buses. Each camera is an IoT node, providing a viewpoint to the cloud for inference. 

We crop the images to the bounding boxes to obtain (potentially) six different viewpoints for each object. We then center and resize each image to $32\times 32\times 3$ pixels, keeping the RGB channels. For views in which a particular object is obscured (and therefore does not exist in the object bounding box list), we generate an empty (black) image~\cite{harvard}. A single data instance is then a collection of six images of an object from the six cameras, and an associated label.

\begin{table}[]
\centering
\scriptsize
\caption{Reliability settings for all experiments.}
\vspace{-5pt}
\begin{tabular}{@{}lcc@{}}
\toprule
\multicolumn{3}{c}{{\bf Reliability Setting $R_V$}} \\ \midrule
Experiment       & Health & Camera \\ 
~ & (order: $[f_1, f_2, e_1]$) & (order: $[f_1, f_2, f_3, f_4, e_1, e_2, e_3, e_4]$) \\ \midrule
 Surviv. Setting & &  \\
~~~{\em Normal}& $[99\%, 98\%, 96\%]$ & $[99.5\%, 99\%, 98\%, 97\%, 95\%, 95\%, 95\%, 95\%]$\\
~~~{\em Poor}           & $[98\%, 96\%, 92\%]$ & $[99\%, 98\%, 94\%, 93\%, 90\%, 90\%, 87\%, 87\%]$\\
~~~{\em Hazardous}              &[$90\%, 85\%, 80\%]$ & $[90\%, 90\%, 80\%, 80\%, 70\%, 60\%, 70\%, 66\%]$\\
\bottomrule
\end{tabular}
\label{tab:surv-configs}
\end{table}

In total, the dataset contains around 1,400 data points. Since the distribution of the classes is skewed towards the ``car'' class, we utilize weighted (i.e. cost-sensitive) cross-entropy loss to incur a higher penalty for incorrectly predicting any given data point to be car \cite{he2008learning}. Although the dataset contains very few images for the bus class, we keep it as one of the classes.

For both experiments, we separate each dataset into train, validation, and test with an 80/10/10 split. We use the validation set to select the best model among different {\em training epochs}. We run the model for many training epochs, select the model with highest validation accuracy, and report its accuracy on the test set, which the DNN has never seen.

\subsection{Experiment Setup}
We implemented our experiments on Google Cloud using TensorFlow and Keras. In order to assess the performance of our proposed method on different architectures of distributed DNNs, we propose separate model configurations for each experiment. For the \health{} experiment (Fig.~\ref{fig:experiment_setup}a), we consider a vertically distributed DNN that consists of ten hidden layers of width 250. The DNN layers are partitioned into four physical nodes (an edge node, two fog nodes, and a cloud node). The edge node contains one layer, the first fog node two, the second fog node three, and the cloud node four hidden layers. Conversely, for the \camera{} experiment (Fig.~\ref{fig:experiment_setup}b) we consider a DNN consisting of 14 layers of width 32 that are distributed both vertically and horizontally. The hidden layers are partitioned into nine physical nodes (four edge nodes, four fog nodes, and a cloud node). Images from the six individual cameras are merged using element-wise addition. We designed this highly distributed DNN architecture for the \camera{} experiment, since it represents a very general architecture, where the DNN is distributed vertically and horizontally and is also asymmetric.  

In our experiments, we only include the skip hyperconnections that skip one physical node, since, in the edge-fog-cloud hierarchy, the number of physical nodes over which the DNN is distributed is not large. Moreover, we experimented with models having skip hyperconnections that skip more than one physical node, but did not observe any performance gain.

Each of the aforementioned models is trained via stochastic gradient descent using the \emph{Adam} optimizer~\cite{kingma2014adam}. Batch sizes of 1024 and 64, and the learning rates of 0.001 and 0.1 are used for the \health{} and \camera{} experiments, respectively. 

We propose three different reliability settings outlined in Table~\ref{tab:surv-configs}: the setting \high{} indicates reasonable network survivabilities while the settings \med{} and \low{} represent reliability settings (only for the sake of our experiments) when the failures are very frequent in the network. We also have the reliability setting {\em No Failure}, in which all survivabilities are simply set to 1. The failure probabilities of fog nodes and edge nodes in both experiments are described in Table~\ref{tab:surv-configs}. Fog nodes are denoted with $f_i$ and edge nodes with $e_i$. We assume that the cloud node is always available and does not fail. If IoT nodes fail, the distributed DNNs will have no input, and random guessing must be performed. Since we are studying the resiliency of distributed DNNs (and not their input), we did not consider the failure of IoT nodes. 

\subsection{Results}

\begin{figure}[t]
    \centering
    \includegraphics[width=0.84\linewidth]{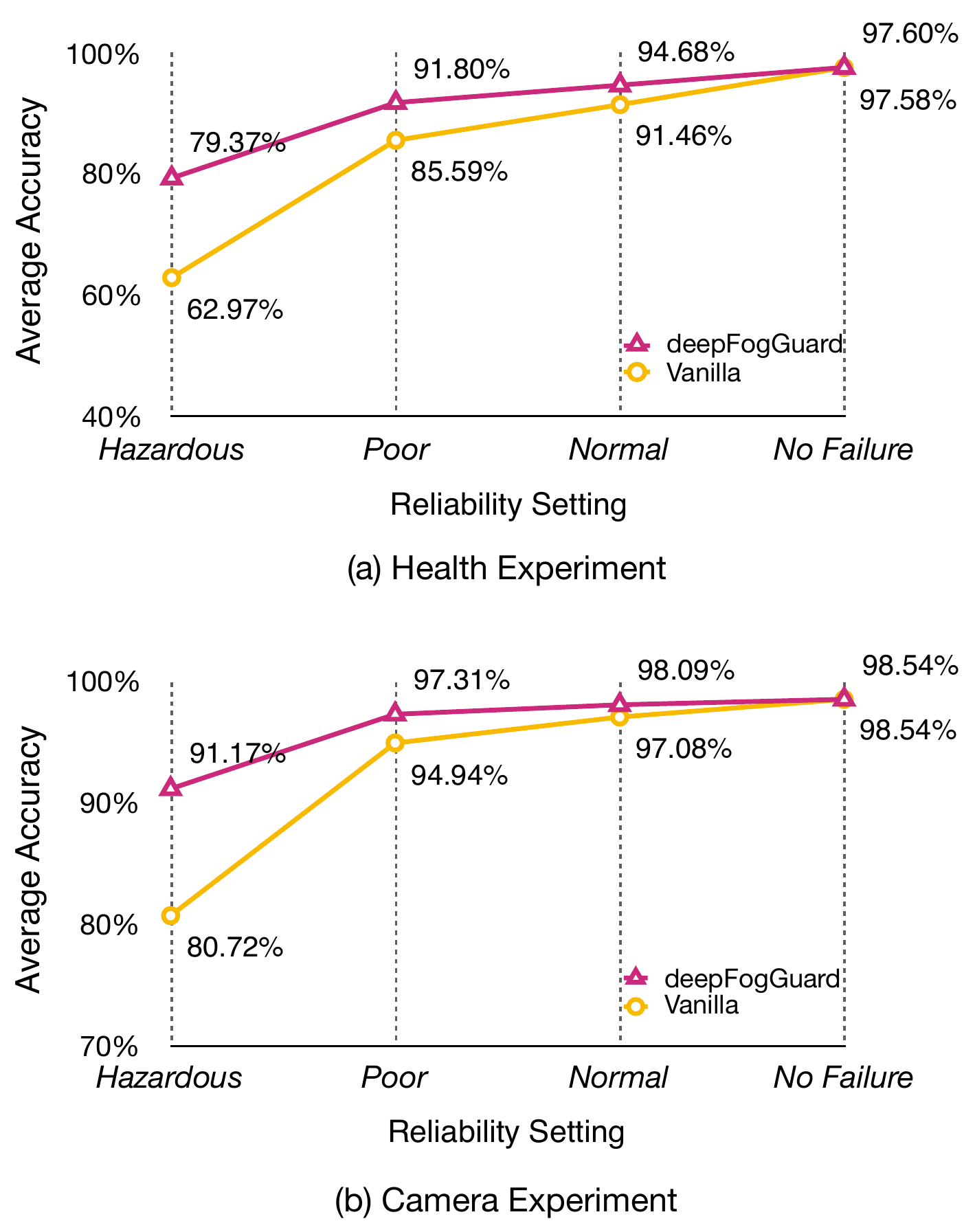}
    \caption{Average accuracy ($\%$) vs. reliability setting}
    \label{fig:main-results}
\end{figure}

Fig.~\ref{fig:main-results} shows the average accuracy of \schemeName{} and Vanilla for each experiment under different reliability settings over 10 runs ($\overline{\mathbb{A}}(G_V, R_V)$). We can see that \schemeName{} is successful in increasing the resiliency of the distributed DNNs, drastically outperforming Vanilla. 

In the \health{} experiment (Fig.~\ref{fig:main-results}a), in the \low{} reliability setting, \schemeName{} increases average accuracy by almost 16\%, relative to the baseline Vanilla. Similarly, in the \med{} reliability setting, the average accuracy of \schemeName{} is around 6\% higher than that of Vanilla. This difference in accuracy decreases when the reliability of the network is higher (e.g. in \high{} or \nofailure{}). Vanilla's poor performance under physical node failure is an indication of the inaccessibility of a path for information flow, hence the occurrence of random guessing for the classification task.

In the \camera{} experiment, we observe the same trends as in the \health{} experiment. Since the architecture of the DNNs is highly distributed (vertically and horizontally) in this experiment, even Vanilla can perform well (above $80\%$). This is because Vanilla's distributed DNN architecture still receives partial data in certain cases of failure. For instance, there are built-in redundancies among cameras, edge nodes, and fog nodes 3 and 4. This is not true for Vanilla in the \health{} experiment, in which the distributed DNN is only vertically partitioned. Similar to the \health{} experiment, we see the largest improvement in the \low{} reliability setting, although \schemeName{} outperforms Vanilla across the board, in both experiments, under various reliability settings.

This concludes the discussion of our experiments. In the next section, we explain the state of the art in this direction, and we position our work's novelty in the literature. Finally, in Section \ref{Conclusion}, we discuss the limitations and opportunities to improve \schemeName{}.

\section{Related Work} \label{related}
The related work in this space can be categorized as follows.

{\bf a. Distributed training}. Training of distributed DNNs has received significant attention from both academia and industry. Some examples include distributed training frameworks from Google \cite{tensorFlow}, Facebook \cite{Pytorch}, Microsoft \cite{microsoft}, and Uber \cite{uber}. Distributed training of DNNs across edge nodes is studied in~\cite{Shiqiang} (non-resilient) and~\cite{pmlr-v80-chen18l, Damaskinos} (resilient against adversaries). Nevertheless, inference in distributed DNNs is less explored. Recently, some IoT application scenarios have emerged that need ongoing and long inference tasks \cite{harvard, morshed2017deep, EdgeEye, hotedge-distributed, ChuangHu, dey1}.  In line with this direction, we study inference of distributed DNNs, but differently, we consider failure resiliency. 

{\bf b. DNN Partitioning}. DNN partitioning frameworks consider several factors to find the best partition map to split and distribute a DNN~\cite{neurosurgeon, wangplacement, dey1, zhou2019distributing, ChuangHu, Li:Edge, elgamal2018droplet}. These frameworks can be used to provide input (partitioned distributed DNN) to \schemeName{}, from which resilient distributed DNN for inference is extracted.

{\bf c. DNN Fault Tolerance}. In the DNN literature, a concept related to failure is {\em fault}, which is when units or weights become defective (i.e. stuck at a certain value, random bit flip, weight fault, or short circuit) \cite{torres2017fault}. Studies on fault tolerance of neural networks date back to the early 90s and are limited to mathematical models with simplistic assumptions (e.g. neural networks with one hidden layer, unit-only and weight-only faults, or sigmoid-only neural networks) \cite{mehrotra1994fault, Bolt92faulttolerance, phatak1995complete}. However, none of these works consider the failure of physical nodes that potentially cause the failure of a large group of DNN units and weights.

{\bf d. DNN Failure Robustness and Resiliency}. Some early works study the resiliency of {\em non-distributed} DNNs against {\em single or multiple} unit failures \cite{zhou2003evolving, sequin1990fault}. More recently, the authors of~\cite{el2017robustness} provide theoretical definitions and bounds for the failure of elements in non-distributed DNNs. Contrary to previous works, this article studies resiliency of {\em distributed} DNN inference in the presence of failure of a {\em large group of DNN units}.

\section{Conclusion} \label{Conclusion}
We presented \schemeName{}, a method for failure resiliency of distributed DNN inference. We confirmed through experiments that skip hyperconnections increase the resiliency of distributed DNNs. \schemeName{} has a few limitations and opportunities for improvement, which we discuss below.

{\bf Limitations}: While \schemeName{} improves resiliency by sending the data along redundant paths, it also inevitably consumes additional bandwidth when there are no failures. Moreover, keeping multiple TCP connections active and checking the status of other physical nodes consumes resources. 

 {\bf Future Work}: This study opens many related research opportunities. Firstly, it is interesting to see how \schemeName{} can be extended to neural networks that have residual connections, or other types of neural networks, such as convolutional neural networks (CNNs) and recurrent neural networks (RNNs). Furthermore, regularization or methods that implicitly increase robustness, such as dropout, may improve the resiliency even more. Finally, one could consider changing the weights of the remaining hyperconnections after the failure of the physical nodes to account for the change in relative input scale of the physical node.

\begin{acks}
We would like to thank professor Vibhav Gogate for the early discussions about resilient DNNs, and Ahmad Darki and professor Keith Winstein for their valuable ideas and comments. 
\end{acks}

\bibliographystyle{ACM-Reference-Format}
\bibliography{Master}


\begin{thebibliography}{00}


\ifx \showCODEN    \undefined \def \showCODEN     #1{\unskip}     \fi
\ifx \showDOI      \undefined \def \showDOI       #1{#1}\fi
\ifx \showISBNx    \undefined \def \showISBNx     #1{\unskip}     \fi
\ifx \showISBNxiii \undefined \def \showISBNxiii  #1{\unskip}     \fi
\ifx \showISSN     \undefined \def \showISSN      #1{\unskip}     \fi
\ifx \showLCCN     \undefined \def \showLCCN      #1{\unskip}     \fi
\ifx \shownote     \undefined \def \shownote      #1{#1}          \fi
\ifx \showarticletitle \undefined \def \showarticletitle #1{#1}   \fi
\ifx \showURL      \undefined \def \showURL       {\relax}        \fi
\providecommand\bibfield[2]{#2}
\providecommand\bibinfo[2]{#2}
\providecommand\natexlab[1]{#1}
\providecommand\showeprint[2][]{arXiv:#2}

\bibitem[\protect\citeauthoryear{Abadi, Barham, Chen, Chen, Davis, Dean, Devin,
  Ghemawat, Irving, Isard, et~al\mbox{.}}{Abadi et~al\mbox{.}}{2016}]%
        {tensorFlow}
\bibfield{author}{\bibinfo{person}{Mart{\'\i}n Abadi}, \bibinfo{person}{Paul
  Barham}, \bibinfo{person}{Jianmin Chen}, \bibinfo{person}{Zhifeng Chen},
  \bibinfo{person}{Andy Davis}, \bibinfo{person}{Jeffrey Dean},
  \bibinfo{person}{Matthieu Devin}, \bibinfo{person}{Sanjay Ghemawat},
  \bibinfo{person}{Geoffrey Irving}, \bibinfo{person}{Michael Isard},
  {et~al\mbox{.}}} \bibinfo{year}{2016}\natexlab{}.
\newblock \showarticletitle{Tensorflow: a system for large-scale machine
  learning.}. In \bibinfo{booktitle}{{\em OSDI}}, Vol.~\bibinfo{volume}{16}.
  \bibinfo{pages}{265--283}.
\newblock


\bibitem[\protect\citeauthoryear{Ali, Anjum, Yaseen, Zamani, Balouek-Thomert,
  Rana, and Parashar}{Ali et~al\mbox{.}}{2018}]%
        {dnnApplication}
\bibfield{author}{\bibinfo{person}{Muhammad Ali}, \bibinfo{person}{Ashiq
  Anjum}, \bibinfo{person}{M~Usman Yaseen}, \bibinfo{person}{A~Reza Zamani},
  \bibinfo{person}{Daniel Balouek-Thomert}, \bibinfo{person}{Omer Rana}, {and}
  \bibinfo{person}{Manish Parashar}.} \bibinfo{year}{2018}\natexlab{}.
\newblock \showarticletitle{Edge enhanced deep learning system for large-scale
  video stream analytics}. In \bibinfo{booktitle}{{\em IEEE 2nd International
  Conference on Fog and Edge Computing (ICFEC)}}. IEEE, \bibinfo{pages}{1--10}.
\newblock


\bibitem[\protect\citeauthoryear{Banos, Villalonga, Garcia, Saez, Damas,
  Holgado-Terriza, Lee, Pomares, and Rojas}{Banos et~al\mbox{.}}{2015}]%
        {mhealth}
\bibfield{author}{\bibinfo{person}{Oresti Banos}, \bibinfo{person}{Claudia
  Villalonga}, \bibinfo{person}{Rafael Garcia}, \bibinfo{person}{Alejandro
  Saez}, \bibinfo{person}{Miguel Damas}, \bibinfo{person}{Juan~A
  Holgado-Terriza}, \bibinfo{person}{Sungyong Lee}, \bibinfo{person}{Hector
  Pomares}, {and} \bibinfo{person}{Ignacio Rojas}.}
  \bibinfo{year}{2015}\natexlab{}.
\newblock \showarticletitle{Design, implementation and validation of a novel
  open framework for agile development of mobile health applications}.
\newblock \bibinfo{journal}{{\em Biomedical engineering online\/}}
  \bibinfo{volume}{14}, \bibinfo{number}{2} (\bibinfo{year}{2015}).
\newblock


\bibitem[\protect\citeauthoryear{Bolt}{Bolt}{1992}]%
        {Bolt92faulttolerance}
\bibfield{author}{\bibinfo{person}{George~Ravuama Bolt}.}
  \bibinfo{year}{1992}\natexlab{}.
\newblock {\em \bibinfo{title}{Fault Tolerance in Artificial Neural Networks}}.
\newblock \bibinfo{thesistype}{Ph.D. Dissertation}. \bibinfo{school}{University
  of York}.
\newblock


\bibitem[\protect\citeauthoryear{Chen, Wang, Charles, and Papailiopoulos}{Chen
  et~al\mbox{.}}{2018}]%
        {pmlr-v80-chen18l}
\bibfield{author}{\bibinfo{person}{Lingjiao Chen}, \bibinfo{person}{Hongyi
  Wang}, \bibinfo{person}{Zachary Charles}, {and} \bibinfo{person}{Dimitris
  Papailiopoulos}.} \bibinfo{year}{2018}\natexlab{}.
\newblock \showarticletitle{{DRACO}: {B}yzantine-resilient Distributed Training
  via Redundant Gradients}. In \bibinfo{booktitle}{{\em Proceedings of the 35th
  International Conference on Machine Learning}}, Vol.~\bibinfo{volume}{80}.
  \bibinfo{publisher}{PMLR}, \bibinfo{pages}{903--912}.
\newblock
\showURL{%
\url{http://proceedings.mlr.press/v80/chen18l.html}}


\bibitem[\protect\citeauthoryear{Chen, Zhao, Li, and Zhao}{Chen
  et~al\mbox{.}}{2019}]%
        {chen2019exploring}
\bibfield{author}{\bibinfo{person}{Yitao Chen}, \bibinfo{person}{Kaiqi Zhao},
  \bibinfo{person}{Baoxin Li}, {and} \bibinfo{person}{Ming Zhao}.}
  \bibinfo{year}{2019}\natexlab{}.
\newblock \showarticletitle{Exploring the Use of Synthetic Gradients for
  Distributed Deep Learning across Cloud and Edge Resources}. In
  \bibinfo{booktitle}{{\em 2nd $\{$USENIX$\}$ Workshop on Hot Topics in Edge
  Computing (HotEdge 19)}}.
\newblock


\bibitem[\protect\citeauthoryear{Chilimbi, Suzue, Apacible, and
  Kalyanaraman}{Chilimbi et~al\mbox{.}}{2014}]%
        {microsoft}
\bibfield{author}{\bibinfo{person}{Trishul~M Chilimbi}, \bibinfo{person}{Yutaka
  Suzue}, \bibinfo{person}{Johnson Apacible}, {and} \bibinfo{person}{Karthik
  Kalyanaraman}.} \bibinfo{year}{2014}\natexlab{}.
\newblock \showarticletitle{Project Adam: Building an Efficient and Scalable
  Deep Learning Training System.}. In \bibinfo{booktitle}{{\em OSDI}},
  Vol.~\bibinfo{volume}{14}. \bibinfo{pages}{571--582}.
\newblock


\bibitem[\protect\citeauthoryear{Damaskinos, El~Mhamdi, Guerraoui, Guirguis,
  and Rouault}{Damaskinos et~al\mbox{.}}{2019}]%
        {Damaskinos}
\bibfield{author}{\bibinfo{person}{Georgios Damaskinos},
  \bibinfo{person}{El~Mahdi El~Mhamdi}, \bibinfo{person}{Rachid Guerraoui},
  \bibinfo{person}{Arsany Hany~Abdelmessih Guirguis}, {and}
  \bibinfo{person}{Sébastien Louis~Alexandre Rouault}.}
  \bibinfo{year}{2019}\natexlab{}.
\newblock \showarticletitle{AGGREGATHOR: Byzantine Machine Learning via Robust
  Gradient Aggregation}.
\newblock  (\bibinfo{year}{2019}).
\newblock
\newblock
\shownote{Conference on Systems and Machine Learning (SysML) 2019, Stanford,
  CA, USA.}


\bibitem[\protect\citeauthoryear{Dey, Mondal, and Mukherjee}{Dey
  et~al\mbox{.}}{2019}]%
        {dey1}
\bibfield{author}{\bibinfo{person}{Swarnava Dey}, \bibinfo{person}{Jayeeta
  Mondal}, {and} \bibinfo{person}{Arijit Mukherjee}.}
  \bibinfo{year}{2019}\natexlab{}.
\newblock \showarticletitle{Offloaded Execution of Deep Learning Inference at
  Edge: Challenges and Insights}. In \bibinfo{booktitle}{{\em 2019 IEEE
  International Conference on Pervasive Computing and Communications Workshops
  (PerCom Workshops)}}. IEEE, \bibinfo{pages}{855--861}.
\newblock


\bibitem[\protect\citeauthoryear{El~Mhamdi, Guerraoui, and Rouault}{El~Mhamdi
  et~al\mbox{.}}{2017}]%
        {el2017robustness}
\bibfield{author}{\bibinfo{person}{EM El~Mhamdi}, \bibinfo{person}{R
  Guerraoui}, {and} \bibinfo{person}{S Rouault}.}
  \bibinfo{year}{2017}\natexlab{}.
\newblock \showarticletitle{On the robustness of a neural network}. In
  \bibinfo{booktitle}{{\em 2017 IEEE 36th Symposium on Reliable Distributed
  Systems (SRDS)}}. \bibinfo{pages}{84--93}.
\newblock


\bibitem[\protect\citeauthoryear{Elgamal, Sandur, Nguyen, Nahrstedt, and
  Agha}{Elgamal et~al\mbox{.}}{2018}]%
        {elgamal2018droplet}
\bibfield{author}{\bibinfo{person}{Tarek Elgamal}, \bibinfo{person}{Atul
  Sandur}, \bibinfo{person}{Phuong Nguyen}, \bibinfo{person}{Klara Nahrstedt},
  {and} \bibinfo{person}{Gul Agha}.} \bibinfo{year}{2018}\natexlab{}.
\newblock \showarticletitle{DROPLET: Distributed Operator Placement for IoT
  Applications Spanning Edge and Cloud Resources}. In \bibinfo{booktitle}{{\em
  2018 IEEE 11th International Conference on Cloud Computing (CLOUD)}}. IEEE,
  \bibinfo{pages}{1--8}.
\newblock


\bibitem[\protect\citeauthoryear{He and Garcia}{He and Garcia}{2008}]%
        {he2008learning}
\bibfield{author}{\bibinfo{person}{Haibo He} {and} \bibinfo{person}{Edwardo~A
  Garcia}.} \bibinfo{year}{2008}\natexlab{}.
\newblock \showarticletitle{Learning from imbalanced data}.
\newblock \bibinfo{journal}{{\em IEEE Transactions on Knowledge \& Data
  Engineering\/}} \bibinfo{number}{9} (\bibinfo{year}{2008}),
  \bibinfo{pages}{1263--1284}.
\newblock


\bibitem[\protect\citeauthoryear{He, Zhang, Ren, and Sun}{He
  et~al\mbox{.}}{2016}]%
        {ms-residual}
\bibfield{author}{\bibinfo{person}{Kaiming He}, \bibinfo{person}{Xiangyu
  Zhang}, \bibinfo{person}{Shaoqing Ren}, {and} \bibinfo{person}{Jian Sun}.}
  \bibinfo{year}{2016}\natexlab{}.
\newblock \showarticletitle{Deep residual learning for image recognition}. In
  \bibinfo{booktitle}{{\em Proceedings of the IEEE conference on computer
  vision and pattern recognition}}. \bibinfo{pages}{770--778}.
\newblock


\bibitem[\protect\citeauthoryear{Hu, Bao, Wang, and Liu}{Hu
  et~al\mbox{.}}{2019}]%
        {ChuangHu}
\bibfield{author}{\bibinfo{person}{Chuang Hu}, \bibinfo{person}{Wei Bao},
  \bibinfo{person}{Dan Wang}, {and} \bibinfo{person}{Fengming Liu}.}
  \bibinfo{year}{2019}\natexlab{}.
\newblock \showarticletitle{Dynamic Adaptive DNN Surgery for Inference
  Acceleration on the Edge}. In \bibinfo{booktitle}{{\em IEEE INFOCOM 2019-IEEE
  Conference on Computer Communications}}. IEEE, \bibinfo{pages}{1423--1431}.
\newblock


\bibitem[\protect\citeauthoryear{Kang, Hauswald, Gao, Rovinski, Mudge, Mars,
  and Tang}{Kang et~al\mbox{.}}{2017}]%
        {neurosurgeon}
\bibfield{author}{\bibinfo{person}{Yiping Kang}, \bibinfo{person}{Johann
  Hauswald}, \bibinfo{person}{Cao Gao}, \bibinfo{person}{Austin Rovinski},
  \bibinfo{person}{Trevor Mudge}, \bibinfo{person}{Jason Mars}, {and}
  \bibinfo{person}{Lingjia Tang}.} \bibinfo{year}{2017}\natexlab{}.
\newblock \showarticletitle{Neurosurgeon: Collaborative intelligence between
  the cloud and mobile edge}. In \bibinfo{booktitle}{{\em ACM SIGARCH Computer
  Architecture News}}, Vol.~\bibinfo{volume}{45}. ACM,
  \bibinfo{pages}{615--629}.
\newblock


\bibitem[\protect\citeauthoryear{Kingma and Ba}{Kingma and Ba}{2014}]%
        {kingma2014adam}
\bibfield{author}{\bibinfo{person}{Diederik~P Kingma} {and}
  \bibinfo{person}{Jimmy Ba}.} \bibinfo{year}{2014}\natexlab{}.
\newblock \showarticletitle{Adam: A method for stochastic optimization}.
\newblock \bibinfo{journal}{{\em arXiv preprint arXiv:1412.6980\/}}
  (\bibinfo{year}{2014}).
\newblock


\bibitem[\protect\citeauthoryear{Li, Zhou, and Chen}{Li et~al\mbox{.}}{2018}]%
        {Li:Edge}
\bibfield{author}{\bibinfo{person}{En Li}, \bibinfo{person}{Zhi Zhou}, {and}
  \bibinfo{person}{Xu Chen}.} \bibinfo{year}{2018}\natexlab{}.
\newblock \showarticletitle{Edge Intelligence: On-Demand Deep Learning Model
  Co-Inference with Device-Edge Synergy}. In \bibinfo{booktitle}{{\em
  Proceedings of the 2018 Workshop on Mobile Edge Communications}} {\em
  (\bibinfo{series}{MECOMM'18})}. \bibinfo{publisher}{ACM},
  \bibinfo{pages}{31--36}.
\newblock


\bibitem[\protect\citeauthoryear{Liu, Qi, and Banerjee}{Liu
  et~al\mbox{.}}{2018}]%
        {EdgeEye}
\bibfield{author}{\bibinfo{person}{Peng Liu}, \bibinfo{person}{Bozhao Qi},
  {and} \bibinfo{person}{Suman Banerjee}.} \bibinfo{year}{2018}\natexlab{}.
\newblock \showarticletitle{EdgeEye: An Edge Service Framework for Real-time
  Intelligent Video Analytics}. In \bibinfo{booktitle}{{\em Proceedings of the
  1st International Workshop on Edge Systems, Analytics and Networking}}. ACM,
  \bibinfo{pages}{1--6}.
\newblock


\bibitem[\protect\citeauthoryear{Mehrotra, Mohan, Ranka, and Chiu}{Mehrotra
  et~al\mbox{.}}{1994}]%
        {mehrotra1994fault}
\bibfield{author}{\bibinfo{person}{Kishan Mehrotra},
  \bibinfo{person}{Chilukuri~K Mohan}, \bibinfo{person}{Sanjay Ranka}, {and}
  \bibinfo{person}{Ching-tai Chiu}.} \bibinfo{year}{1994}\natexlab{}.
\newblock \bibinfo{booktitle}{{\em Fault tolerance of neural networks}}.
\newblock \bibinfo{type}{{T}echnical {R}eport}. \bibinfo{institution}{Syracuse
  University}.
\newblock
\newblock
\shownote{Tech. Rep. RL-TR-94-93. Syracuse University.}


\bibitem[\protect\citeauthoryear{Morshed, Jayaraman, Sellis, Georgakopoulos,
  Villari, and Ranjan}{Morshed et~al\mbox{.}}{2017}]%
        {morshed2017deep}
\bibfield{author}{\bibinfo{person}{Ahsan Morshed},
  \bibinfo{person}{Prem~Prakash Jayaraman}, \bibinfo{person}{Timos Sellis},
  \bibinfo{person}{Dimitrios Georgakopoulos}, \bibinfo{person}{Massimo
  Villari}, {and} \bibinfo{person}{Rajiv Ranjan}.}
  \bibinfo{year}{2017}\natexlab{}.
\newblock \showarticletitle{Deep osmosis: Holistic distributed deep learning in
  osmotic computing}.
\newblock \bibinfo{journal}{{\em IEEE Cloud Computing\/}} \bibinfo{volume}{4},
  \bibinfo{number}{6} (\bibinfo{year}{2017}), \bibinfo{pages}{22--32}.
\newblock


\bibitem[\protect\citeauthoryear{Park, Samarakoon, Bennis, and Debbah}{Park
  et~al\mbox{.}}{2018}]%
        {park2018wireless}
\bibfield{author}{\bibinfo{person}{Jihong Park}, \bibinfo{person}{Sumudu
  Samarakoon}, \bibinfo{person}{Mehdi Bennis}, {and}
  \bibinfo{person}{M{\'e}rouane Debbah}.} \bibinfo{year}{2018}\natexlab{}.
\newblock \showarticletitle{Wireless network intelligence at the edge}.
\newblock \bibinfo{journal}{{\em arXiv preprint arXiv:1812.02858\/}}
  (\bibinfo{year}{2018}).
\newblock


\bibitem[\protect\citeauthoryear{Paszke, Gross, Chintala, and Chanan}{Paszke
  et~al\mbox{.}}{2017}]%
        {Pytorch}
\bibfield{author}{\bibinfo{person}{Adam Paszke}, \bibinfo{person}{Sam Gross},
  \bibinfo{person}{Soumith Chintala}, {and} \bibinfo{person}{Gregory Chanan}.}
  \bibinfo{year}{2017}\natexlab{}.
\newblock \showarticletitle{Pytorch: Tensors and dynamic neural networks in
  python with strong gpu acceleration}.
\newblock  (\bibinfo{year}{2017}).
\newblock


\bibitem[\protect\citeauthoryear{Phatak and Koren}{Phatak and Koren}{1995}]%
        {phatak1995complete}
\bibfield{author}{\bibinfo{person}{Dhananjay~S Phatak} {and}
  \bibinfo{person}{Israel Koren}.} \bibinfo{year}{1995}\natexlab{}.
\newblock \showarticletitle{Complete and partial fault tolerance of feedforward
  neural nets}.
\newblock \bibinfo{journal}{{\em IEEE Transactions on Neural Networks\/}}
  \bibinfo{volume}{6}, \bibinfo{number}{2} (\bibinfo{year}{1995}),
  \bibinfo{pages}{446--456}.
\newblock


\bibitem[\protect\citeauthoryear{{Roig}, {Boix}, {Ben Shitrit}, and
  {Fua}}{{Roig} et~al\mbox{.}}{2011}]%
        {multiview-dataset}
\bibfield{author}{\bibinfo{person}{G. {Roig}}, \bibinfo{person}{X. {Boix}},
  \bibinfo{person}{H. {Ben Shitrit}}, {and} \bibinfo{person}{P. {Fua}}.}
  \bibinfo{year}{2011}\natexlab{}.
\newblock \showarticletitle{Conditional Random Fields for multi-camera object
  detection}. In \bibinfo{booktitle}{{\em 2011 International Conference on
  Computer Vision}}. \bibinfo{pages}{563--570}.
\newblock
\showISSN{2380-7504}


\bibitem[\protect\citeauthoryear{Sequin and Clay}{Sequin and Clay}{1990}]%
        {sequin1990fault}
\bibfield{author}{\bibinfo{person}{Carlo~H Sequin} {and} \bibinfo{person}{RD
  Clay}.} \bibinfo{year}{1990}\natexlab{}.
\newblock \showarticletitle{Fault tolerance in artificial neural networks}. In
  \bibinfo{booktitle}{{\em 1990 IJCNN international joint conference on neural
  networks}}. IEEE, \bibinfo{pages}{703--708}.
\newblock


\bibitem[\protect\citeauthoryear{Sergeev and Balso}{Sergeev and Balso}{2018}]%
        {uber}
\bibfield{author}{\bibinfo{person}{Alexander Sergeev} {and}
  \bibinfo{person}{Mike~Del Balso}.} \bibinfo{year}{2018}\natexlab{}.
\newblock \showarticletitle{Horovod: fast and easy distributed deep learning in
  {TensorFlow}}.
\newblock \bibinfo{journal}{{\em arXiv preprint arXiv:1802.05799\/}}
  (\bibinfo{year}{2018}).
\newblock


\bibitem[\protect\citeauthoryear{Srivastava, Greff, and Schmidhuber}{Srivastava
  et~al\mbox{.}}{2015}]%
        {highway}
\bibfield{author}{\bibinfo{person}{Rupesh~K Srivastava}, \bibinfo{person}{Klaus
  Greff}, {and} \bibinfo{person}{J{\"u}rgen Schmidhuber}.}
  \bibinfo{year}{2015}\natexlab{}.
\newblock \showarticletitle{Training very deep networks}. In
  \bibinfo{booktitle}{{\em Advances in neural information processing systems
  (NeurIPS)}}. \bibinfo{pages}{2377--2385}.
\newblock


\bibitem[\protect\citeauthoryear{Tao and Li}{Tao and Li}{2018}]%
        {hotedge-distributed}
\bibfield{author}{\bibinfo{person}{Zeyi Tao} {and} \bibinfo{person}{Qun Li}.}
  \bibinfo{year}{2018}\natexlab{}.
\newblock \showarticletitle{eSGD: Communication Efficient Distributed Deep
  Learning on the Edge}. In \bibinfo{booktitle}{{\em {USENIX} Workshop on Hot
  Topics in Edge Computing (HotEdge 18)}}. \bibinfo{publisher}{{USENIX}
  Association}, \bibinfo{address}{Boston, MA}.
\newblock
\showURL{%
\url{https://www.usenix.org/conference/hotedge18/presentation/tao}}


\bibitem[\protect\citeauthoryear{Teerapittayanon, McDanel, and
  Kung}{Teerapittayanon et~al\mbox{.}}{2017}]%
        {harvard}
\bibfield{author}{\bibinfo{person}{Surat Teerapittayanon},
  \bibinfo{person}{Bradley McDanel}, {and} \bibinfo{person}{HT Kung}.}
  \bibinfo{year}{2017}\natexlab{}.
\newblock \showarticletitle{Distributed deep neural networks over the cloud,
  the edge and end devices}. In \bibinfo{booktitle}{{\em Distributed Computing
  Systems (ICDCS), 2017 IEEE 37th International Conference on}}. IEEE,
  \bibinfo{pages}{328--339}.
\newblock


\bibitem[\protect\citeauthoryear{Torres-Huitzil and Girau}{Torres-Huitzil and
  Girau}{2017}]%
        {torres2017fault}
\bibfield{author}{\bibinfo{person}{Cesar Torres-Huitzil} {and}
  \bibinfo{person}{Bernard Girau}.} \bibinfo{year}{2017}\natexlab{}.
\newblock \showarticletitle{Fault and error tolerance in neural networks: A
  review}.
\newblock \bibinfo{journal}{{\em IEEE Access\/}}  \bibinfo{volume}{5}
  (\bibinfo{year}{2017}), \bibinfo{pages}{17322--17341}.
\newblock


\bibitem[\protect\citeauthoryear{Wang, Tuor, Salonidis, Leung, Makaya, He, and
  Chan}{Wang et~al\mbox{.}}{2019}]%
        {Shiqiang}
\bibfield{author}{\bibinfo{person}{Shiqiang Wang}, \bibinfo{person}{Tiffany
  Tuor}, \bibinfo{person}{Theodoros Salonidis}, \bibinfo{person}{Kin~K Leung},
  \bibinfo{person}{Christian Makaya}, \bibinfo{person}{Ting He}, {and}
  \bibinfo{person}{Kevin Chan}.} \bibinfo{year}{2019}\natexlab{}.
\newblock \showarticletitle{Adaptive federated learning in resource constrained
  edge computing systems}.
\newblock \bibinfo{journal}{{\em IEEE Journal on Selected Areas in
  Communications\/}} \bibinfo{volume}{37}, \bibinfo{number}{6}
  (\bibinfo{year}{2019}), \bibinfo{pages}{1205--1221}.
\newblock


\bibitem[\protect\citeauthoryear{Wang, Zafer, and Leung}{Wang
  et~al\mbox{.}}{2017}]%
        {wangplacement}
\bibfield{author}{\bibinfo{person}{Shiqiang Wang}, \bibinfo{person}{Murtaza
  Zafer}, {and} \bibinfo{person}{Kin~K Leung}.}
  \bibinfo{year}{2017}\natexlab{}.
\newblock \showarticletitle{Online placement of multi-component applications in
  edge computing environments}.
\newblock \bibinfo{journal}{{\em IEEE Access\/}}  \bibinfo{volume}{5}
  (\bibinfo{year}{2017}), \bibinfo{pages}{2514--2533}.
\newblock


\bibitem[\protect\citeauthoryear{Yao, Zhao, Zhang, Su, and Abdelzaher}{Yao
  et~al\mbox{.}}{2017}]%
        {Deepiot}
\bibfield{author}{\bibinfo{person}{Shuochao Yao}, \bibinfo{person}{Yiran Zhao},
  \bibinfo{person}{Aston Zhang}, \bibinfo{person}{Lu Su}, {and}
  \bibinfo{person}{Tarek Abdelzaher}.} \bibinfo{year}{2017}\natexlab{}.
\newblock \showarticletitle{Deepiot: Compressing deep neural network structures
  for sensing systems with a compressor-critic framework}. In
  \bibinfo{booktitle}{{\em Proceedings of the 15th ACM Conference on Embedded
  Network Sensor Systems}}. ACM, \bibinfo{pages}{4}.
\newblock


\bibitem[\protect\citeauthoryear{Yousefpour, Fung, Nguyen, Kadiyala, Jalali,
  Niakanlahiji, Kong, and Jue}{Yousefpour et~al\mbox{.}}{2019}]%
        {fog-survey}
\bibfield{author}{\bibinfo{person}{Ashkan Yousefpour}, \bibinfo{person}{Caleb
  Fung}, \bibinfo{person}{Tam Nguyen}, \bibinfo{person}{Krishna Kadiyala},
  \bibinfo{person}{Fatemeh Jalali}, \bibinfo{person}{Amirreza Niakanlahiji},
  \bibinfo{person}{Jian Kong}, {and} \bibinfo{person}{Jason~P Jue}.}
  \bibinfo{year}{2019}\natexlab{}.
\newblock \showarticletitle{All one needs to know about fog computing and
  related edge computing paradigms: A complete survey}.
\newblock \bibinfo{journal}{{\em Journal of Systems Architecture\/}}
  \bibinfo{volume}{98} (\bibinfo{year}{2019}), \bibinfo{pages}{289 -- 330}.
\newblock


\bibitem[\protect\citeauthoryear{Zhou, Wen, Teodorescu, and Du}{Zhou
  et~al\mbox{.}}{2019}]%
        {zhou2019distributing}
\bibfield{author}{\bibinfo{person}{Li Zhou}, \bibinfo{person}{Hao Wen},
  \bibinfo{person}{Radu Teodorescu}, {and} \bibinfo{person}{David~HC Du}.}
  \bibinfo{year}{2019}\natexlab{}.
\newblock \showarticletitle{Distributing Deep Neural Networks with
  Containerized Partitions at the Edge}. In \bibinfo{booktitle}{{\em 2nd
  $\{$USENIX$\}$ Workshop on Hot Topics in Edge Computing (HotEdge 19)}}.
\newblock


\bibitem[\protect\citeauthoryear{Zhou and Chen}{Zhou and Chen}{2003}]%
        {zhou2003evolving}
\bibfield{author}{\bibinfo{person}{Zhi-Hua Zhou} {and} \bibinfo{person}{Shi-Fu
  Chen}.} \bibinfo{year}{2003}\natexlab{}.
\newblock \showarticletitle{Evolving fault-tolerant neural networks}.
\newblock \bibinfo{journal}{{\em Neural Computing \& Applications\/}}
  \bibinfo{volume}{11}, \bibinfo{number}{3-4} (\bibinfo{year}{2003}),
  \bibinfo{pages}{156--160}.
\newblock


\end{thebibliography}


%
%
%
%
%
%
%
%

\end{document}